\documentclass[12pt]{article}
\newcommand{\be}{\begin{equation}}
\newcommand{\ee}{\end{equation}}
\newcommand{\beas}{\begin{eqnarray*}}
\newcommand{\eeas}{\end{eqnarray*}}
\newcommand{\bea}{\begin{eqnarray}}
\newcommand{\eea}{\end{eqnarray}}
\newcommand{\ba}{\begin{array}}
\newcommand{\ea}{\end{array}}
\newcommand{\nn}{\nonumber}

\newcommand{\de}{\delta}
\newcommand{\la}{\lambda}
\newcommand{\si}{\sigma} 
\newcommand{\n}{\nabla}

\begin{document}
\title{
{\bf 
Minimal metagravity vs.\  dark matter and/or dark energy
}}
\author{Yu.\ F.\ Pirogov
\\
\it Theory Division, 
Institute for High Energy Physics,  Protvino, \\
\it RU-142281 Moscow Region, Russia
}
\date{}
\maketitle
\abstract{\noindent 
The minimal metagravity theory, explicitly violating the general
covariance but preserving the unimodular one, is applied to study the
evolution of the isotropic homogeneous Universe. 
The massive scalar graviton, contained in  the theory in addition to
the massless tensor one, is treated as a source of the 
dark matter and/or dark energy. The modified Friedmann equation for
the scale factor of the Universe is derived. The question whether
the minimal metagravity can emulate the LCDM concordance model, valid
in General Relativity, is discussed.\\

}


\section{Introduction}

According to the present-day cosmological paradigm, given by
General Relativity~(GR) and  the standard cosmology, the reasonable
description  of our Universe in toto is achieved in the so-called
LCDM concordance model (for a review of cosmology, 
see, e.g.\ ref.~\cite{astro}). 
In accordance with the model, the Universe is spatially flat, fairly
isotropic and homogeneous being filled predominantly with the dark
energy, accounted for by the $\Lambda$-term, as well as with the cold
dark matter in the energy proportion  roughly ${3:1}$. The energy
fraction of the luminous matter is almost negligible. The  nature of
the dark energy and the dark matter seems to be the main puzzle of the
contemporary physics. 

Thus, all the sources of the dark substances, including the
indirect ones,  are to be investigated.
With this in mind,  we study in the present paper whether the above
substances (or the parts of them) can be mimicked by a modification of
GR, namely, the minimal metagravity theory
proposed earlier~\cite{Pir1}.\footnote{For a brief exposition of the
theory, see, ref.~\cite{Pir2}.}  Due to the explicit violation of the
general covariance (GC) to the unimodular covariance (UC),  such a
theory describes the massive scalar graviton in
addition to the massless tensor one. The idea is to try and
associate the scalar graviton with the dark matter and/or dark energy. 
In Section~2, the compendium of the minimal metagravity theory  is
given. In Section~3, the evolution of the
isotropic homogeneous Universe in the framework of such a theory  is
considered, and the modified Friedmann equation for the scale factor
of the Universe is derived. It is argued then in Resume that the
minimal metagravity is not explicitly inconsistent with the LCDM
concordance model motivating thus for the further study.

\section{Minimal metagravity}

To begin with, let us present the short compendium of the metagravity
theory. Under the latter, we  understand generally the effective field
theory of the metric  revealing, due to the explicit  GC violation,
the extra physical degrees of freedom contained in the
metric  besides those for 
the massless tensor graviton. In the case of the minimal violation, to
be used in what follows, the metagravity preserves the residual
UC and describes for this reason only one
additional particle, the massive scalar graviton. The generic action
of such a minimal metagravity is as follows
\be\label{GCV}
S=\int\Big(L_{\rm g}(g_{\mu\nu})+  L_{\si}(g_{\mu\nu},\si)
+ L_{\rm m}( \phi_{\rm m}, g_{\mu\nu},\si) \Big) \sqrt{-g}\,d^4x,
\ee
where $g_{\mu\nu}$ is the dynamical metric, $\phi_{\rm m}$ is the
generic  matter field and 
\be
\si=\frac{1}{2}\ln\frac{g}{\tilde g}.
\ee 
In the above, $g=\mbox{\rm det\,}g_{\mu\nu}$ and  $\tilde g$ is an
absolute (nondynamical) scalar density of the same  weight as~$g$. 
Depending on the ratio of the two similar scalar densities, $\si$ is
the scalar and thus can serve as the Lagrangian field variable. This
field corresponds to the metric compression waves and is to be treated
as representing the scalar graviton. 
  
In eq.~(\ref{GCV}), $L_{\rm g}$ is the generally covariant Lagrangian
for the tensor graviton being chosen conventionally in the extended
Einstein-Hilbert form:
\be\label{EH}
L_{\rm g}=- \frac{1}{2} m_{\rm P}^2  R(g_{\mu\nu}) +\Lambda,
\ee
where $R=g^{\mu\nu} R_{\mu\nu}$ is the Ricci scalar, with 
$R_{\mu\nu}$ being the Ricci curvature, and $\Lambda$ is the
cosmological constant. Also, $m_{\rm P}=(8\pi G_{\rm N})^{-1/2}$ is
the Planck mass, with $G_{\rm N}$ being the Newton's constant. $
L_\si$ is the scalar graviton  Lagrangian  looking in the lowest order
on the derivatives as follows
\be\label{Ls}
L_\si= \frac{1}{2} f_\si^2\,\partial
\si\cdot\partial \si-V_\si(\si).
\ee
Here $f_\si$ is a constant with the dimension of mass and $ V_\si$
is the potential producing the mass for the scalar graviton.
The dependence of $\si$  on  $\tilde g$ explicitly violates
the part of GC, namely the local scale covariance,
still  preserving the residual UC. A priori, one expects 
$f_\si ={\cal O}(m_{\rm P})$. Also, one  expects that $ V_\si$  though
being  nonzero  is suppressed. Finally, $L_{\rm m}$ is the matter
Lagrangian possessing, generally, only the residual UC, too. 

Varying the action~(\ref{GCV}), with respect to $g^{\mu\nu}$,
$\tilde g$ being fixed, one arrives at the minimal metagravity
equation:
\be\label{eomg}
G_{\mu\nu} = \frac{1}{m_{\rm P}^2}\Big( T^{(\rm m)}_{\mu\nu} + 
T^{(\si)}_{\mu\nu}+ T^{(\Lambda)}_{\mu\nu} \Big),
\ee
with   
\be
G_{\mu\nu}=R_{\mu\nu}-\frac{1}{2} R g_{\mu\nu}
\ee
being the gravity tensor. The r.h.s.\ of eq.~(\ref{eomg}) is  the
total energy-momentum of the nontensor-graviton origin, produced by
the matter and  the scalar gravitons, plus the
vacuum energy. $T^{(\rm m)}_{\mu\nu}$ is the  matter
energy-momentum tensor including, if required, the real dark matter
contribution, too. For the matter as the continuous medium, of
interest in cosmology, one has 
\be
T^{(\rm m)}_{\mu\nu}=(\rho_{\rm m}+p_{\rm m})u_\mu u_\nu -p_{\rm m}
g_{\mu\nu},
\ee
with  $\rho_{\rm m}$ being the energy
density, $p_{\rm m}$  the pressure and $u^\mu$ the medium
4-velocity. $T^{(\si)}_{\mu\nu}$ is the scalar graviton  contribution: 
\be\label{DT}
T^{(\si)}_{\mu\nu}=
f_\si^2 \Big(\partial_\mu\si \partial_\nu\si-
\frac{1}{2}\partial\si\cdot \partial\si\, g_{\mu\nu} \Big) 
+V_{\si} g_{\mu\nu} +\Big(f_\si^2\n\cdot\n \si+
V'_{\si}\Big) g_{\mu\nu},
\ee
with $V'_\si =  \partial V_{\si}/\partial \si$, the covariant
derivative $\n_\mu \si=\partial_\mu \si$ and 
\be
\n\cdot\n \si=\frac{1}{\sqrt {-g}} 
\partial_\mu(\sqrt {-g}g^{\mu\nu}\partial_\nu\si). 
\ee
The term $T^{(\si)}_{\mu\nu}$ can
be treated as the scalar graviton contribution to the
dark matter and/or dark energy. Finally, 
\be
T^{(\Lambda)}_{\mu\nu}=-p_\Lambda g_{\mu\nu}, 
\ee
with
$\rho_\Lambda+p_\Lambda =0$ and  $\rho_\Lambda=
m_{\rm P}^2\Lambda$,
is the vacuum contribution to the dark energy. Under
$\Lambda>0$, it produces the negative pressure.
Due to the Bianchi identity, $\nabla_\mu G^{\mu\nu}=0$,  and the
property $\nabla_\la g_{\mu\nu}=0$, the  energy-momentum  of the
matter and the scalar gravitons is conserved covariantly: 
\be\label{cc}
\nabla_\mu (T_{\rm m}^{\mu\nu} +  T_{\si}^{\mu\nu} )=0,
\ee
whereas  the energy-momentum of the matter alone,
$T^{(\rm m)}_{\mu\nu}$, ceases to conserve.

\section{Modified Friedmann equation}

In the properly chosen observer's coordinates
$x^\mu=(t, \rho, \theta, \varphi )$  the
Friedmann-Robertson-Walker solution for the interval in the isotropic
homogeneous Universe is
\be\label{ds}
ds^2=dt^2-a^2(t)\Big(
\frac{1}{ 1-\kappa^2 \rho^2}\,d \rho^2+
\rho^2(d \theta^2+\sin^2\theta d\varphi^2)\Big),
\ee
with $\kappa$ being a constant with the
dimension of mass. This interval is form-invariant relative to the
shift of the origin  of the spatial coordinates, reflecting the
isotropy and  homogeneity of the Universe.    
Conventionally, one can rescale  the unit of mass so that
$\kappa^2 = k$, with $k=\pm1,0$. The last three cases correspond,
respectively, to the spatially closed, open and  flat Universe. In
eq.~(\ref{ds}), the spatial factor $1/ ( 1-\kappa^2 \rho^2)$ is the
geometrical one, given a priori, while the temporal scale factor
$a(t)$ is the dynamical one to be determined by the gravity equations.

Choosing the new radial variable $r$
\be
\rho= \frac{r}{1+\kappa^2 r^2/4}
\ee
one gets
\be
ds^2=dt^2-\frac{a^2(t)}{ (1+\kappa^2 r^2/4)^2}\,d{\bf x}^2,
\ee
with $x^\mu=(x^0= t, \{x^m\}= {\bf x})$, $m=1,2,3$ 
and $r^2= {\bf x}^2$. In other terms, the metric looks like
\bea
g_{00}&=&1, \ \ g_{m0}=0,\nn\\
g_{mn}&=&-\frac{a^{2}(t)}{(1+\kappa^2 r^2/4)^2}\,\de_{mn},
\eea
with  $\sqrt {-g}=a^{3}/(1+\kappa^2 r^2/4)^{3}$.
These coordinates will be understood in what follows. 

From eq.~(\ref{DT}), one gets 
\be
T^{(\si)}_{m0}=f_\si^2\partial_m \si \dot\si,
\ee
where $\dot \si=d \si/d t$ and
\be
\si=\ln \frac{ a^{3}(t)}{\sqrt{- \tilde g}\,(1+\kappa^2 r^2/4)^3   }.
\ee
Hence, generally, $T^{(\si)}_{m0}\neq 0$.
On the other hand due to the isotropy, there should fulfil
$G_{m 0}=0$ and $T^{(\rm m)}_{m0}=0$.  This requires
$T^{(\si)}_{m0}=0$, too, and thus $\partial_m \si=0$. To achieve this
the spatial parts of $ g$ and $\tilde g$ should coincide, so that
\be
\sqrt{- \tilde g}=\frac{ \tilde a^{3}(t)}{(1+\kappa^2 r^2/4)^3},
\ee
with  $\tilde a(t)$ being a free temporal factor. Altogether one has
\be
\si(t)=3\ln \frac{a(t)}{\tilde a(t)}.
\ee

The minimal metagravity equation~(\ref{eomg}) results now in the two
following equations 
\bea\label{MG}
G^0_0&=&3 \left(\Big(\frac{\dot a}{a}\Big)^2+\frac{\kappa^2}{a^2}
\right)=\frac{1}{m_{\rm
P}^2}
\rho,\nn\\
G_n^m&=&\left(2\, \frac{\ddot a}{a}
+\Big(\frac{\dot a}{a}\Big)^2 + \frac{\kappa^2}{a^2} \right)\de^m_n
=-\frac{1}{m_{\rm P}^2}  p\, \de^m_n ,
\eea
with $\ddot a =d^2 a/d t^2$, and $\rho$ and $p$ being
the total energy density and pressure, respectively:
\bea
\rho&=&\rho_{\rm m}+\rho_\si +\rho_\Lambda,\nn\\
p&=&p_{\rm m}+p_\si+p_\Lambda 
\eea
(remind that $p_\Lambda=-\rho_\Lambda$).
The continuous medium is taken to be nonrelativistic:
$u^0=1$, $u^m=0$. The energy density and pressure for the
scalar gravitons are formally defined as for the continuous medium:  
\bea\label{SG}
T_{{\si}}{}_0^0 &=&  
f_\si^2\Big( \frac{1}{2}\dot\si^2  +3 \frac{\dot
a}{a}\dot\si+\ddot\si\Big )+(V_\si+V'_\si)\equiv\rho_\si ,\nn\\
T_{{\si}}{}^m_n &=&  
\bigg[f_\si^2\Big(- \frac{1}{2}\dot\si^2  +3 \frac{\dot
a}{a}\dot\si+\ddot\si \Big)+(V_\si+V'_\si)\bigg]\de^m_n  \equiv
-p_\si \de^m_n, 
\eea
with the effective ``equation of state''
\be\label{SG'}
\rho_\si+p_\si=f_\si^2 \dot\si^2.
\ee
Note that $\rho_\si$ and $p_\si$ are coordinate dependent, in
distinction with the scalars $\rho_{\rm m}$ and $p_{\rm m}$. In the
above, one has
\bea\label{SG''}
\dot\si&=&3\Big( \frac{\dot a}{a}-\tilde H\Big),\nn\\
\ddot\si&=& 3\bigg( \frac{\ddot a}{a} -\Big(\frac{\dot
a}{a}\Big)^2-\dot{\tilde H}\bigg),
\eea
with $\tilde H\equiv \dot{\tilde a}/\tilde a$,  and use is made of the
relation 
\be
\n\cdot\n\si=\ddot\si+3 \frac{\dot a}{a}\dot\si.
\ee

With account for equations~(\ref{SG}) and (\ref{SG''}), the two lines
of eq.~(\ref{MG}) substitute the similar equations valid in GR. 
The first line of eq.~(\ref{MG}), the modified Friedmann equation,
determines  the scale factor $a(t)$, with the second line giving the
consistency condition. Introducing the ``critical'' energy density 
\be
\rho_{\rm c}=3 m_{\rm P}^2\Big(\frac{\dot a}{a}\Big)^2
\ee
one can bring the modified Friedmann equation to the form
\be\label{F1}
\Omega\equiv \frac{\rho}{\rho_{\rm c}}=1+  
\frac{\kappa^2}{{\dot a}^2}.
\ee
Further, differentiating the first line of eq.~(\ref{MG}) and
combining the result with the second line, one can substitute
the latter by the continuity condition:
\be\label{F2}
\dot\rho+ 3\frac{\dot a}{a}(\rho+p)=0,
\ee
with $\rho_\si+p_\si =f_\si^2\dot\si^2$ and  
$\rho_\Lambda+p_\Lambda =0$.

To really solve these equations one should specify the free  functions
entering the theory. For the continuous medium, there are
conventionally two extreme cases: the dust
\be
\rho_{\rm m}=\frac{\rho_0}{a^3}, \ \ p_{\rm m}=0
\ee
and the radiation
\be
\rho_{\rm m}=3p_{\rm m}=\frac{\rho_0}{a^4}.
\ee
For the scalar graviton, the good starting point would conceivably be
the assumption $\tilde a={\rm Const}$ and thus $\tilde H=0$.  As for
the potential $V_\si$, little can be said about it
a priori,  and probably it should be looked for by the trial-and-error
method. With these caveats, the  equations above
are ready for use in working out the cosmological scenarios in the
framework of the minimal metagravity.

\section{Resume}

The above system of the modified Friedmann equation plus the
continuity condition is much more intricate compared with the similar
system valid in GR and urge to a special investigation. Nevertheless,
to show that the given approach is not completely unrealistic suppose
that there indeed exists  such a solution which mimics
the observationally consistent  GR solution for the spatially flat
Universe ($\kappa^2=0$). The latter solution corresponds to the 
LCDM concordance model and requires the existence of the cold dark
matter ($p_{\rm dm}=0$) in the energy fraction 
$\rho_{\rm dm}/\rho_{\rm c}$ roughly $1/4$. For this
to be reconciled in the minimal metagravity, one should get
effectively for the scalar graviton $p_\si\simeq0$ and thus,
according to eq.~(\ref{SG'}),  $\rho_\si\simeq
f^2_\si\dot\si^2 $. Now, putting $\rho_\si/\rho_{\rm c}\simeq 1/4$, as
observations imply, and assuming $\tilde H=0$ one gets $f_\si\simeq
m_{\rm P}/2\sqrt 3$ as the necessary condition. The relation  
$f_\si ={\cal O}(m_{\rm P})$  appears to be
quite natural. Similarly, one can look for the minimal metagravity
solutions which mimic the (part of) dark energy together with the
cold dark matter. Whether such an approach can really be made
consistent with observations remains yet to be seen.

\end{document}